\documentclass[12pt]{article}
\usepackage{amsfonts}
\newsavebox{\uuunit}
\sbox{\uuunit}
    {\setlength{\unitlength}{0.825em}
     \begin{picture}(0.6,0.7)
        \thinlines
        \put(0,0){\line(1,0){0.5}}
        \put(0.15,0){\line(0,1){0.7}}
        \put(0.35,0){\line(0,1){0.8}}
       \multiput(0.3,0.8)(-0.04,-0.02){12}{\rule{0.5pt}{0.5pt}}
     \end {picture}}

\begin{document}
 %%%%%%%%%%%%%%%%%%%%%%%%%%%%%%%%%%%%%%%%%%%%%%%%%%%%%%%%%%%
%\begin{titlepage}
\begin{flushright}
FTUV-07-0110 \hskip .4cm IFIC/07-10
\\ Jan. 10, 2007
\end{flushright}
\vspace{.5cm}

\begin{center}
 {\bf \large Twistor string as tensionless superstring}\footnote{
 Talk delivered at the Workshop of the RTN network {\it Constituents, fundamental forces
 and Symmetries of the Universe}, Napoli October 9-13, 2006, to
 appear in the proceedings (Fortschritte der Physik).}

\bigskip

{ Igor A. Bandos$^{\dagger,\,\ddagger}$, Jos\'e A. de
Azc\'arraga$^\dagger$, C\`esar Miquel-Espanya$^\dagger$}

\bigskip

%Put the addresses here
$^\dagger$ {\it Department of Theoretical Physics, Valencia
University, and IFIC (CSIC-UVEG), 46100-Burjassot (Valencia),
Spain}

$^\ddagger${\small\it Institute for Theoretical Physics, NSC KIPT,
 61108 Kharkov, Ukraine}

\end{center}

\bigskip

\begin{abstract}
We give a brief review of the twistor string approach to
supersymmetric Yang-Mills theories with an emphasis on the
different formulations of (super)string models in supertwistor
space and their superspace form. We discuss the classical
equivalence among the Siegel closed twistor string and the Lorentz
harmonics formulation of the ($N=4$) tensionless superstring, and
notice the possible relation of the twistor string to the $D=10$
Green-Schwarz superstring action, as well as to models in the
enlarged, tensorial superspaces that are relevant in higher spin
theories.
\end{abstract}

\section{MHV Yang-Mills amplitudes, twistors and supertwistors}

At the end of 2003, E. Witten looked again \cite{Witten03} at the
connection between N=4 supersymmetric Yang-Mills (SYM) theory and
string theory using the twistor approach \cite{Pen}. In contrast
to the conventional AdS/CFT correspondence \cite{AdS}, the above
connection establishes a link between the weak coupling limits on
both sides and, hence, it can be checked perturbatively. This led
to the development of a new technique to compute gauge theory
amplitudes \cite{CSW} (the Cachazo-Svr\v{c}ek-Witten [CSW] or MHV
diagram rules, see below) and renewed the interest in the Penrose
twistor program \cite{Pen} of replacing spacetime by twistors.
\medskip

 {\it 1.1. MHV amplitudes through bosonic spinors}

The amplitude for the scattering of $n$ gauge bosons with 2
positive and ($n$-$2$) negative helicities reads
\begin{equation}
\label{AmIJ}
A(1,2,...,n) = 2^{n/2} ig^{n-2} Tr(t^{a_1}... t^{a_n})
{<IJ>^4 \over <12><23> \ldots <n1> }\;  \qquad \end{equation} (see
\cite{Nair88,MHV2006}and refs. therein), where
%$2^{\frac{n}{2}}$ is a numerical factor,
$g$ is the YM coupling constant, $Tr(t^{a_1},\dots,t^{a_n})$ is
the trace of the product of $n$ gauge group generators, and the
one-particle matrix elements $<12>, ... , < n1>$ are expressed
through the contraction of pairs of bosonic spinors, a spinor for
each gluon state, {\it e.g.},
\begin{equation}
\label{<12>}
<12> = \lambda^{\alpha 1} \lambda^2_{\alpha}= \epsilon^{\alpha
\beta} \lambda^{1}_\beta \lambda^2_{\alpha}\equiv -
\lambda^{\alpha 2} \lambda^1_{\alpha}= -<21> \; .
\end{equation}
Specifically, $I$ and $J$ in (\ref{AmIJ}) ($A(1,2,...,n)  \propto
<IJ>^4$) refer to two positive helicity gluons, while the gluons
with $n\neq I,J$ are assumed to have negative helicity. Notice
that the maximally helicity violating or MHV amplitude
(\ref{AmIJ}) is holomorphic: it depends on the $\lambda$'s through
their contractions $<12>$, etc. In contrast, the complex conjugate
spinors $\bar{\lambda}^1_{\dot{\alpha}}:=
({\lambda}^1_{{\alpha}})^*$, $\ldots$,
$\bar{\lambda}^n_{\dot{\alpha}}:= ({\lambda}^n_{{\alpha}})^*$,
{\it etc.},  the contractions of which are denoted by $[1,2]:=
\bar{\lambda}^{1\dot{\alpha}} \bar{\lambda}^2_{\dot{\alpha}}\equiv
-[2,1]$, are not present in (\ref{AmIJ}).

One may ask, why the $n$ particle amplitudes can be expressed in
terms of just $n$ bosonic spinors $\lambda^i_\alpha$? The answer
is that a bosonic spinor can be used to describe the {\it
on-shell} momentum and the helicity that characterize (modulo
internal quantum numbers) the on-shell state of a massless
particle. The momentum of such particle is lightlike, $p^2:=p_\mu
p^\mu=0$, a condition that is solved in spinor space by \cite{Pen}
\begin{eqnarray}\label{p=lbl}
p_{\alpha\dot{\alpha}}:= p_\mu
\sigma^\mu_{\alpha\dot{\alpha}}=\lambda_{\alpha
}\bar{\lambda}_{\dot{\alpha}}\quad & \Leftrightarrow \quad p_\mu=
{1\over 2} \lambda\sigma^\mu\bar{\lambda} := {1\over 2}
{\sigma}_{\mu \alpha\dot{\alpha}}\lambda^{\alpha
}\bar{\lambda}^{\dot{\alpha}} \; .
\end{eqnarray}
The light-likeness of the vector $p_\mu$ given by the Penrose
formula (\ref{p=lbl}) follows from $p^2\delta_\alpha^\beta=
p_{\alpha\dot\alpha}\tilde p^{\dot\alpha\beta}= \lambda_\alpha
\bar{\lambda}_{\dot{\alpha}} \bar{\lambda}^{\dot{\alpha}}
\lambda^\beta$,  which vanishes identically.

The polarization vectors $\varepsilon^{\mu \pm}$ of the positive
and negative helicity particles are defined to be lightlike and
orthogonal to the lightlike momentum, $(\varepsilon^{\pm})^2=0$,
$\varepsilon^{\mu \pm} p_\mu=0$. They are expressed by
\begin{eqnarray}
\label{e=lu}   \varepsilon^{-i}_{\alpha\dot{\alpha}} =
{\lambda_{\alpha}{}^{\!\!\! i}\, \bar{u}_{\dot{\alpha}} \;/
[\bar{\lambda}^{i}\!\! ,\,\bar{u}] }\quad , \qquad {} \qquad
\varepsilon^{+i}_{\alpha\dot{\alpha}} =
{u_{\alpha}\bar{\lambda}_{\dot{\alpha}}{}^{\!\!\!\! i}\; /
 <u,\, {\lambda}^{i}> }\quad,
\end{eqnarray}
using a constant reference spinor $\bar u_{\dot\alpha}
(u_\alpha=(\bar u_{\dot\alpha})^* )$. The relativistic invariance
condition corresponds to the requirement that the observable
quantities are independent of the choice of $u$.

Thus, all the kinematical information on the {\it on-shell} state
of a gauge boson, its helicity and its lightlike momentum, are
encoded in a single bosonic spinor. This is the reason why the
scattering amplitude for $n$ gauge bosons can be condensed in an
expression written in terms of bosonic spinors, as (\ref{AmIJ})
above.

\medskip

{\it 1.2. The Cachazo-Svr\v{c}ek-Witten or MHV diagram technique
inspired by the twistor string}

The CSW diagram technique \cite{CSW} consists in cutting a Feynman
diagram into MHV pieces (which is always possible \cite{CSW}) and
then treating them as vertices connected by scalar propagators.
Clearly, there is an immediate problem: by cutting a Feynman
diagram into MHV pieces, one gets generally subdiagrams in which
one or more legs correspond to virtual particles, {\it i.e.}
particles that are off-shell and for which the basic Penrose
representation (\ref{p=lbl}) does not hold. The prescription
proposed in \cite{CSW} and checked to be true inside and beyond
\cite{MHV2006} the domain of the original $N$=4 supersymmetric
Yang-Mills context, is to associate to a virtual particle the
bosonic spinor defined by
$\lambda_\alpha(p)=p_{\alpha\dot\alpha}\bar\omega^{\dot\alpha}$,
where $\bar\omega^{\dot\alpha}$ is an arbitrary reference spinor.
Relativistic invariance then requires that the amplitude $A$ is
independent of the choice of the reference spinor, ${\partial
A}/{\partial \bar\omega}=0$ . This condition has been shown to
hold for tree and one-loop diagrams in $N$=4 SYM theory, and
checked for some two-loop and some non- and less supersymmetric
theories \cite{MHV2006}.

The equivalence of the MHV and the Feynman diagram calculus was
originally proved for the tree diagrams of $N$=4 SYM theory
\cite{CSW}. It was then extended to one-loop diagrams \cite{CSW}
and also checked for some higher-loop ones as well as for less
supersymmetric ($N$=2, $N$=1 and non-supersymmetric $N$=0) YM
theories  (see \cite{MHV2006} for a recent review). However, the
original version was developed for $N$=4 SYM theories and was
inspired by the twistor string model \cite{Witten03}, which is a
string model formulated in the space of $N$=4 supertwistors. The
action principles which lie beyond the twistor string
\cite{Witten03,NB04,Siegel04} and its spacetime (superspace)
formulation will be our main subject here.

To begin, we briefly address two questions: 1) what is a twistor?
and 2) what is a supertwistor?

\medskip

{\it 1.3 Penrose twistors and Ferber supertwistors}

A twistor \cite{Pen} can be understood as a Dirac spinor. It has
four complex components in two Weyl spinors,
$\Upsilon^{\hat{\alpha}} = (\lambda_{\alpha} ,
\mu^{\dot{\alpha}})\, \in\, \mathbb{C}^4$, and provides a
spinorial representation for the conformal group $SO(2, 4)$ as
well as a fundamental representation for the (locally isomorphic)
$SU(2,2)=Spin(2,4)$ group.

Twistor space is usually considered as a complex projective space,
\begin{eqnarray}
\label{Y=CP}
 \Upsilon^{\hat{\alpha}} \sim z \Upsilon^{\hat{\alpha}} \quad \Rightarrow \quad
\Upsilon^{\hat{\alpha}}= (\lambda_\alpha \; ,
\mu^{\dot{\alpha}})\; \in\; \mathbb{CP}^3\; \quad .
\end{eqnarray}
The reason for the identification $\Upsilon^{\hat{\alpha}} \sim z
\Upsilon^{\hat{\alpha}}$ of twistors that differ by  a complex
factor $z\in \mathbb{C}\backslash \{0\}$ is the obvious complex
scale invariance of the Penrose incidence relation
\begin{eqnarray}\label{mu=xl}\mu^{\dot{\alpha}} = x^{\dot{\alpha}\alpha}
\lambda_\alpha\; , \qquad x^{\dot{\alpha}\alpha}:= x^\mu
\tilde{\sigma}_\mu^{\dot{\alpha}\alpha}\; , \quad
\end{eqnarray}
which defines a spacetime point $x^\mu$  or, more precisely, a
lightlike line in Minkowski space,
$\hat{x}^{\dot{\alpha}\alpha}(\tau)= x^{\dot{\alpha}\alpha}  +
\tau  \, \lambda^\alpha \bar{\lambda}^{\dot{\alpha}}$. Eq.
(\ref{mu=xl}) is the general solution for the (`helicity')
constraint $ \bar{\Upsilon}_{\hat{\alpha}} \,
{\Upsilon}^{\hat{\alpha}}  := \bar{\lambda}_{\dot{\alpha}} \,
{\mu}^{\dot{\alpha}} - \bar{\mu}^{{\alpha}} {\lambda}_{{\alpha}}
=0$.

  Ferber {\it supertwistors} \cite{FS78-83}, $\Upsilon^{\Sigma}:=
(\Upsilon^{\hat{\alpha}}\, , \eta_i) = (\lambda_\alpha\, ,
\mu^{\dot{\alpha}} \, , \eta_i)\; \in\; \mathbb{C}^{(4|N)}$ ,
($i=1,\dots,N$), carry a fundamental representation of
$SU(2,2|N)$. They  include $N$ fermionic variables $\eta_i$,
$i=1,\dots,N$, in addition to the Penrose twistor
$\Upsilon^{\hat\alpha}\;$. Under the $\Upsilon^{\Sigma} \sim
z\Upsilon^{\Sigma}$ equivalence relation, supertwistors become
homogeneous coordinates of the complex projective superspace
$\mathbb{CP}^{(3|N)}$,
\begin{eqnarray}\label{sTwistor} & \Upsilon^{\Sigma}:= (\Upsilon^{\hat{\alpha}}\, ,
\eta_i) = (\lambda_\alpha\, , \mu^{\dot{\alpha}} \, , \eta_i)\;
\in\; \mathbb{CP}^{(3|N)}\; , \qquad \eta_i\eta_j=-\eta_j\eta_i\;
, \quad i=1,\ldots, N \; .
\end{eqnarray}
The scaling $\Upsilon^{\Sigma} \sim  z \Upsilon^{\Sigma}$ appears
now as a symmetry of the Penrose-Ferber incidence relations,
\begin{eqnarray}\label{mu=eta=} \mu^{\dot{\alpha}} =
x_L^{\dot{\alpha}\alpha} \lambda_\alpha\; , \quad \eta_i=
\theta^\alpha_i \lambda_\alpha \; \qquad
x_L^{\dot{\alpha}\alpha}:= x_L^\mu
\tilde{\sigma}_\mu^{\dot{\alpha}\alpha}:= x^{\dot{\alpha}\alpha} +
2i \theta^\alpha_i \bar{\theta}^{\dot{\alpha}i}\; .
\end{eqnarray}
These involve the coordinates $Z^M:=(x^\mu\, , \,
\theta^\alpha_i\, , \, \bar{\theta}^{\dot{\alpha}i})$ of
$N$-extended $D$=4 superspace and define a $(1|N)$-dimensional
subsuperspace $\mathbb{R}^{(1|N)}$,
\begin{eqnarray}
 \label{R(1,N)}
 \hat{x}^{\dot{\alpha}\alpha}= x^{\dot{\alpha}\alpha}  + \tau  \, \lambda^\alpha
\bar{\lambda}^{\dot{\alpha}}+2[i\kappa_i{\bar\theta}{}^{{\dot\alpha}i}\lambda^\alpha
+ c.c.] \; , \quad \hat{\theta}_i^\alpha = {\theta}_i^\alpha  +
\kappa_i \lambda^\alpha\; , \quad \quad \{ (\tau\, , \kappa^i) \}
= \mathbb{R}^{(1|N)}\quad,
 \end{eqnarray}
where the $\kappa_i$ are $N$ fermionic parameters. This
$\mathbb{R}^{(1|N)}$ is the Sorokin-Tkach-Volkov-Zheltukhin
worldline superspace \cite{STVZ} (or {\it superworldline}, first
introduced in the context of the spinning superparticle
\cite{spinsup}), the simplest example of the superworldvolumes of
the superembedding approach to superbranes \cite{bpstv,Dima}.

\section{(Super)twistor string action(s)}

The basic worldsheet fields of the twistor string models are the
supertwistors (\ref{sTwistor}). At present there are three main
versions of the twistor string action (see \cite{IB+JdA+C=2006}
for further discussion and references): (i) the constrained
$\mathbb{CP}^{(3|4)}$ sigma model by Witten \cite{Witten03}; (ii)
the open string model by Berkovits \cite{NB04} involving {\it two}
supertwistors; and (iii) the simplest one, proposed by Siegel in
\cite{Siegel04}, described by the closed string action
\begin{eqnarray}\label{TWS-S}
S &=\int_{W^2} e^{++} \wedge \bar{\Upsilon}_\Sigma\;
\nabla\Upsilon^\Sigma \; + d^2\xi L_{G} =  \int d^2 \xi \, [
\sqrt{|\gamma(\xi) |} \,\; \bar{\Upsilon}_\Sigma(\xi)\,
\nabla_{\!_{--}} \Upsilon^\Sigma(\xi) \;  + L_{G} ] \; .
\end{eqnarray}
Here $\bar{\Upsilon}_\Sigma :=
 \left(\Upsilon^\Pi\right)^\ast \Omega_{\Pi\Sigma} = (
\bar{\lambda}_{\dot{\alpha}} \,  , \, - \bar{\mu}^\alpha \,  ; \,
2i \bar{\eta}^i )$ is the $SU(2,2|N)$-adjoint of
$\Upsilon^\Sigma$, $e^{\pm\pm}=d\xi^m e^{\pm\pm}_m(\xi)$ are the
worldsheet zweibein one-forms and $e^{++}\wedge e^{--}=d^2\xi
\sqrt{|\gamma|}$ is the invariant surface element of the string
worldsheet $W^2$. The covariant derivative
$\nabla=e^{++}\nabla_{++}+e^{--}\nabla_{--}=d-iB$ involves the
$U(1)$-connection $B$, which serves as a Lagrange multiplier for
the constraint
\begin{eqnarray}\label{YY=0}
\bar{\Upsilon}_\Sigma\, {\Upsilon}^\Sigma =
\bar{\lambda}_{\dot{\alpha}} \, {\mu}^{\dot{\alpha}} -
\bar{\mu}^{{\alpha}} {\lambda}_{{\alpha}} +2i \bar{\eta}^i \eta_i
=0 \; .
\end{eqnarray}
Finally, $L_G$ in (\ref{TWS-S}) is the lagrangian for the
worldsheet fields used to construct the Yang-Mills symmetry
current. As noted in \cite{NB04}, one can use {\it e.g.}, the
worldsheet fermionic fields $\Psi^I$ in the fundamental
representation of the gauge group. Then, $d^2\xi \, L_{G} =
{1\over 2} e^{++} \wedge (\bar{\psi}_{I} d\psi^I - d\bar{\psi}_{I}
\, \psi^I)$ in the notation of \cite{IB+JdA+C=2006}.

The lagrangian of the open string model (ii) by Berkovits
\cite{NB04} is given by $S = \int_{W^2} e^{++} \wedge
\bar{\Upsilon}^-_{\Sigma } \nabla (\Upsilon^{-\Sigma} ) - e^{--}
\wedge \bar{\Upsilon}^+_{\Sigma} \nabla (\Upsilon^{+ \Sigma} )   +
\int_{W^2} d^2\xi (L^L_{G} + L^R_{G})$. It contains two
supertwistors, one left-moving $\Upsilon^{-\Sigma}$  and one
right-moving $\Upsilon^{+\Sigma}$ , and also two copies of the `YM
current' degrees of freedom, which are `glued' by  boundary
conditions on $\partial W^2$ . The lagrangian form integrated over
the open worldsheet $W^2$ is actually the sum of Siegel's
lagrangian in (\ref{TWS-S}) and its right-moving counterpart.
Finally, the original action for the $\mathbb{CP}^{(3|4)}$ twistor
string model (i) by Witten \cite{Witten03}, expressed in the
present notation, can be found in \cite{IB+JdA+C=2006}.

\section{The twistor string as a tensionless superstring}
As it was shown in \cite{IB+JdA+C=2006}, an equivalent form of the
action (\ref{TWS-S}) is given by the tensionless superstring
action from \cite{BZnull,BZnullr} (called twistor-like
Lorentz-harmonics formulation of the null superstring for reasons
explained in \cite{IB+JdA+C=2006}). This means that, ignoring its
YM part, the action (\ref{TWS-S}) can be written in $D=4$, $N=4$
superspace as
\begin{eqnarray}\label{S-0S}
&S = \int\limits_{W^2} e^{++} \wedge
\hat{\Pi}^{\dot{\alpha}\alpha}
 \bar{\lambda}_{\dot{\alpha}}\lambda_{\alpha}  =
\int\limits_{W^2} e^{++} \wedge (dx^{\dot{\alpha}\alpha} - i
d\theta_i^{\alpha} \bar{\theta}^{\dot{\alpha}i} + i
\theta_i^{\alpha} d\bar{\theta}^{\dot{\alpha}i})
\bar{\lambda}_{\dot{\alpha}}\lambda_{\alpha} \; ,
\end{eqnarray}
where $\hat{\Pi}^{\dot{\alpha}\alpha}
%\equiv d\xi^m\Pi_m^{\dot{\alpha}\alpha} \equiv
= d\tau \Pi_\tau^{\dot{\alpha}\alpha} + d\sigma
\Pi_\sigma^{\dot{\alpha}\alpha}$ is the pull-back to the
worldsheet $W^2$ of the flat supervielbein on  $D=4$, $N=4$
superspace,
 $\Pi^{\dot{\alpha}\alpha} := dx^{\dot{\alpha}\alpha}
- i d\theta_i^{\alpha} \bar{\theta}^{\dot{\alpha}i} + i
\theta_i^{\alpha} d\bar{\theta}^{\dot{\alpha}i}$. This action
possesses an irreducible $\kappa$--symmetry
\begin{eqnarray}\label{kappaS}
&\delta_\kappa x^{\dot{\alpha}\alpha}= i \delta_\kappa
\theta^\alpha_i \, \bar{\theta}^{\dot{\alpha}i} - i
\theta^\alpha_i
 \delta_\kappa \bar{\theta}^{\dot{\alpha}i}  \; ,
 \quad \delta_\kappa \theta^\alpha_i \,= \kappa_i
 \lambda^{\alpha}\; , \quad \delta_\kappa
\bar{\theta}^{\dot{\alpha}i} \, = \bar{\kappa}^{i}
\bar{\lambda}^{\dot{\alpha}} \; , \quad \\
&\delta_\kappa \lambda^{\alpha}= \delta_\kappa
\bar{\lambda}^{\dot{\alpha}}= \delta_\kappa
e^{++}=0\quad.\nonumber\end{eqnarray} This is obtained from the
infinitely reducible  $\kappa$-symmetry \cite{dA+L82+S82}, with
$\delta_\kappa \theta^\alpha_i \, = \kappa_{\dot{\alpha}i}
\Pi_{\!_{--}}^{\dot{\alpha}\alpha}$,
$\delta_\kappa\bar\theta^{\dot\alpha
i}=\Pi_{--}^{\dot\alpha\alpha}\bar\kappa_{\alpha}^{i}$, by using
the relation $
\Pi^{\dot\alpha\alpha}_{--}\sim\bar\lambda^{\dot\alpha}\lambda^\alpha
$ which provides the general solution of the equations of motion
for the bosonic sponor field $\lambda$ (which is an auxiliary
field in the action (\ref{S-0S})). The  $\kappa$-symmetry reduces
the number of degrees of freedom to the same 8+(16/2) of the
twistor string (\ref{TWS-S}).

The simplest way to check the equivalence \cite{IB+JdA+C=2006} of
(\ref{S-0S}) to the first, supertwistor part of the Siegel twistor
string action (\ref{TWS-S}), is to use Leibniz's rule to move the
derivative to act on
%the bosonic spinors
$\lambda$ and to take into account that the Penrose-Ferber
incidence relations (\ref{mu=eta=}) provide the general solution
of the constraint (\ref{YY=0}).

The fact that the twistor string is tensionless \cite{Siegel04}
can be understood by observing the conformal invariance of the
action (\ref{TWS-S}) or (\ref{S-0S}), which implies the absence of
any dimensionful parameters in it. The Berkovits open twistor
string model (ii) is equivalent to the tensionless superstring
moving in the direct sum of two copies of the $D=4$, $N=4$
superspace \cite{IB+JdA+C=2006}.

\section{On the tensionful parent of the twistor string}
The tensionless nature of the twistor string was first noticed by
Siegel \cite{Siegel04}, who also posed the problem of searching
for a possible tensionful parent. Its existence can also be
understood \cite{IB+JdA+C=2006} as a consequence of the results in
\cite{BZnull,BZnullr} according to which the mass spectrum of the
intrinsically tensionless or null string is continuous, while the
also tensionless quantum twistor string
\cite{Witten03,NB04,Siegel04} is assumed to describe the
Yang-Mills theory amplitudes and, hence, must have massless fields
in the quantum state spectrum. In fact, since in the conformal
algebra the dilatation operator does not commute with the square
of the momentum operator, a continuous mass spectrum or a
zero-masses one are the only alternatives for a conformally
invariant theory. The quantization \cite{Lindstrom03+} of the
tensionless superstring, which leads to massless fields in the
spectrum, requires the (explicit or implicit) use of stringy
oscillators, which are the suitable variables for a tensionful
string. In contrast, those of the null-string are rather the
spacetime coordinates and momenta. As a result, the quantization
of the twistor string should correspond to the quantization of the
tensionless limit of a tensionful superstring, rather than that of
the intrinsically tensionless, or null, superstring.

In ref. \cite{Siegel04}, Siegel discussed the possible tensionful
parents of the twistor superstring in a purely bosonic context,
proposing a tensionful QCD string \cite{Sie96} as its bosonic
part. The inclusion of fermions brings in new questions. In
particular, as far as we assume that the tensionless limit has to
be a smooth one,  the number of degrees of freedom should not
change in this limit and, thus, the number of gauge symmetries
should be the same, including the number of fermionic
$\kappa$-symmetries already mentioned.

A detailed discussion of these questions can be found in
\cite{IB+JdA+C=2006}. In short, if we were interested just in the
$N=1,2$ counterparts of the tensionless twistor string, their
tensionful counterparts would be the $D=4$, $N=1,2$ Green-Schwarz
superstrings, as can be seen in the framework of the spinor moving
frame or Lorentz harmonics formulation \cite{BZstr}. In the more
interesting $N=4$ case, the tensionful parent action requires an
extension of the bosonic sector of superspace so that, to obtain
the twistor string, the tensionless limit has to be accompanied by
dimensional reduction. One could conjecture that the tensionful
parent of the twistor string is given by the $D=10$ Green-Schwarz
superstring. To describe such a relation in a simple way one would
have to use the spinor moving frame or the Lorentz harmonics
formulation of the $D=10$ Green-Schwarz superstring
\cite{BZstr,BZstrH}.

\section{Concluding remarks}

We would like to mention that, at the present level of
understanding, the Green-Schwarz superstring does not appear as
the only possible candidate for a tensionful parent of the twistor
string. As discussed in \cite{IB+JdA+C=2006}, one can also
consider supersymmetric string models in enlarged tensorial
superspaces \cite{TnesSSP}, which have found applications in
higher spin theories \cite{BLS99}.

This is not the only possible link between twistor string and
higher spin theories. According to \cite{Lindstrom03+}, the
quantization of a tensionless limit of the Green-Schwarz
superstring should result in a higher spin theory. On the other
hand, as a result of our identification of the twistor string with
the zero-tension superstring, the twistor string description of
the Yang-Mills amplitudes \cite{Witten03,NB04} should also be
related to the quantization of a tensionless superstring (a
tensionless limit of a superstring). It would be interesting to
understand the interrelations and differences between these two
quantizations of tensionless superstring(s) in some detail.

\bigskip
{\bf Acknowledgements}. Partial support from the Spanish Ministry
of Education and Science (FIS2005-02761) and EU FEDER funds, the
Generalitat Valenciana, the Ukrainian State Fund for Fundamental
Research (383), the INTAS project 2006-7928 and the EU `Forces
Universe' network MRTN-CT-2004-005104, is gratefully acknowledged.

\end{document}